\documentclass[]{aa}
\usepackage{graphics}

\begin{document}
\thesaurus{03 (12.03.3;12.12.1)}

\title{The NTT SUSI DEEP FIELD
 \thanks{Based on observations collected at the NTT 3.5m of ESO, Chile}}

\author{S. Arnouts  \inst{1,2}  \and 
        S. D'Odorico \inst{2} \and
        S. Cristiani \inst{1} \and
        S. Zaggia     \inst{2,4} \and
        A. Fontana   \inst{3} \and
        E. Giallongo \inst{3} }
\offprints{S. Arnouts}
\institute{  
  Dipartimento di Astronomia, Universit\`a di Padova, vicolo
dell'Osservatorio 5, I-35122, Padova, Italy
\and
  European Southern Observatory, Karl Schwarzschild Strasse 2,
D-85748 Garching, Germany
\and
  Osservatorio Astronomico di Roma, via dell'Osservatorio, I-00040
Monteporzio, Italy 
\and
  Osservatorio Astronomico di Capodimonte, via Moiariello 15, I-80131 
Napoli, Italy}
\date{Received 15 March 1998}

  \maketitle

  \begin{abstract}

We present a deep BVrI multicolor catalog of galaxies in a 
5.62 sq.arcmin field 80 arcsec south of the high redshift ($z=4.7$) 
quasar BR 1202-0725, derived from observations with the direct CCD 
camera SUSI at the ESO NTT.
The formal 5$\sigma$ magnitude limits (in $2\times$ FWHM apertures) are 
 26.9, 26.5, 25.9  and 25.3 in B, V, r and I respectively.
  Counts, colors for the star and galaxy samples are discussed and
  a comparison with a deep HST image in the I band is presented.
 The percentage of merged or blended galaxies in the SUSI data to this 
 magnitude limit is estimated to be not higher than 1\%.
 
At the same galactic latitude of the HDF but pointing toward the galactic 
center, the star density in this field is found to be $\sim3$ times
higher, with $\sim20\%$ of the objects with V-I$>$3.0. 
Reliable colors have been measured for galaxies selected
down to $r=26$.  The choice of the optical filters has been optimized to
define a robust multicolor selection of galaxies at $3.8\leq z\leq
4.4$.  Within this interval the surface density of galaxy candidates
with $r < 26$ in this field is  $2.7\pm 0.4$ arcmin$^{-2}$ corresponding
to a  comoving density of Star Formation Rate
at $3.8\leq z\leq 4.4$ of $10^{-2.00} \hbox{  -  } 
 10^{-1.82} h^{3}\ M_{\odot} \cdot yr^{-1} \cdot Mpc^{-3}$.

\end{abstract}

\section{Introduction}
 With the implementation of efficient CCD cameras at 4m class-telescopes
 deep imaging in high galactic latitude fields has become a powerful 
 tool to study galaxy evolution. Since the pioneering work of 
 Tyson (1988),
 deep ground -based observations have been extended to different color 
 bands (Metcalfe et al., 1995; Smail et al. 1995, Hogg et al. 1997) with image
quality of 1 arcsec FWHM or better.

 A new benchmark for deep survey work has been set by the Hubble Deep Field 
observations (Williams et al., 1996). Although the size of the HST primary mirror is 2.5m only
and the CCDs in the WFPC2 instrument have relatively poor blue and UV
 sensitivities, the combination of very long integration time, low sky 
 background and sub-arcsec angular size for most of the faint galaxies in the
 field, has led to limiting magnitudes which are more than a factor of ten 
 fainter than the deepest ground-based surveys. \\
  The HST observations,  beside their intrinsic scientific value, 
 acted as a very efficient catalyst for complementary photometric work in
 other bands, notably the IR, and for spectroscopic observations of galaxies
  down  to $m_{I}$=25 with the Keck telescope.  The project also demonstrated the 
 scientific advantage of dedicating a sizable chunk of observing time to
 the deep exploration of a single, size-limited field.

The other crucial development in this field has been the identification by the
"Lyman break" technique (Steidel and Hamilton, 1993 ; Steidel, Pettini
 and Hamilton, 1995) and subsequent spectroscopic follow-up of a large
 number (a few hundreds known at the beginning of 1998) of $z\sim3$  
galaxies with the Keck telescope (Steidel et al. 1996, Steidel et al. 1998).
 These results have permitted to address for the first time the issues 
of star formation rate and clustering at these redshifts, extending the
 previous spectroscopic survey work (Lilly et al. 1995; Cowie, Hu and 
Songaila 1995) beyond z= 1.5 .

 The Lyman break technique is an example (so far the one with the highest 
 success rate) of the photometric redshift techniques.
 Since accurate photometry can be obtained for objects at least
 two magnitudes fainter than the spectroscopic  limit, photometric redshifts,
 that is redshifts which are obtained by comparing broad band observations 
 of galaxies with a library of observed templates or with stellar population
 synthesis models, are 
 the only practicable way to extend the studies of the population of galaxies
 at high redshifts ($z\simeq$4 and beyond) to luminosities below $L_{*}$.
 Photometric redshifts have been derived from ground-based observations 
 to relatively bright magnitude limits and redshifts $z<1$ (Koo 1985,
 Connolly et al. 1995).  
 The data set of the HDF, which goes much deeper, has revived the interest in
 this type of work with significant results (Sawicki et al 1997, Connolly 
 et al. 1997 and references therein to earlier work ). 
 Giallongo et al. (1998) have used ground -based observations with the SUSI CCD camera at the 
 ESO New Technology Telescope to measure photometric redshifts for
  $\sim200$ galaxies down to a limiting magnitude of $m_{r}=25$.

 The observations presented in this paper were obtained for the program 
 "Faint Galaxies in an ultra-deep multicolour SUSI field",
 P.I. S.D'Odorico,
 approved for ESO 
 Period 58 and executed in service mode also at the ESO NTT in February through
 April 1997 in photometric nights with seeing better than 1 arcsec. The
scientific goals were the study of the photometric redshift distribution of
 the faint galaxies and of gravitational shearing in the field. 
The field of view of the SUSI CCD camera is comparable to the HDF, and the 
goal was to reach limiting magnitudes in the four bands which would enable
 photometric redshift estimates to $r_{AB} \sim 26.5$ or about 1.5 magnitude
 fainter than in the previous work by Giallongo et al.(1998).
The final coadded calibrated frames have been made
 available since January 98 at the following address:\\
{\tt http://www.eso.org/research/sci-prog/ndf/}.

The chosen field, hereafter referred to as NTT Deep Field or NTTDF, 
is at 80 arcsec south of the $z=4.7$ QSO BR1202-072 (McMahon et al 1994). 
It is partially overlapping with the field centered on the QSO and studied 
in the same 4 optical bands and in K band by Giallongo et al. (1998).
The high redshift QSO at the center of the field has several known metal 
systems in its line of sight spanning from $z=1.75$ to $z=4.7$ 
(Wampler et al. 1996, Lu et al. 1996) making this field quite interesting
for a future comparison between the absorbers and the properties and
distribution in  redshifts of the field galaxies.

In this paper, we describe the observations, the reduction procedures and the 
objects catalogue in Section 2, and the galaxy counts and colors in Section 3. 
The data in the I band are compared with a deep HST observation of the same 
field and in the same band and more in general with the Hubble Deep Field 
results in Section 4. The selection of high redshift galaxy candidates is 
discussed in Section 5 and the conclusions are presented in Section 6.

In a forthcoming paper (Fontana et al. 1998, to be submitted) the data are 
combined with infrared observations  and used to derive the 
photometric  redshifts of the galaxies in the field.

\section{The Data Sample} 
\subsection {The Observations}
 The field  centered on $\alpha = 12^h \ 05^m \ 22.4^s$, $\delta = -7^{\circ} 
\  44' \  12''$ was observed in service mode with the SUSI imaging CCD camera
 at the Nasmyth focus of the ESO New Technology Telescope.
 The observations have been obtained in four broad band filters, 
 the standard BV passbands of the Johnson-Kron-Cousins system (JKC) and 
  r and i of the Thuan-Gunn system. The CCD was a 1k x 1k thinned, 
anti-reflection coated device (ESO CCD no. 42). The scale on the detector is 
0.13 arcsec/pixel. \\
 The data, including photometric calibrations with standard stars from 
 Landolt (1992), were obtained in the period between February and April 1997.
 In  Table~\ref{tobs}, we summarize 
 the observations and list the expected magnitude limits at 5-sigma within
 apertures of 2$\times$FWHM of the combined frames. 
 \begin{table*}	
\caption{Log of photometric observations} \label{tobs}
\begin{tabular}{cccccccc}
\hline
Filter &  Number of  &  total &  seeing range& Final PSF & Zero-point & $\mu(1\sigma)$  & Mag. limit \\
       & frames & exp. time (s) &(arcsec)& (arcsec)  & $^{(1)}$ &  mag/arcsec$^2$ & (5 $\sigma$) $^{(1,2)}$ \\
\hline
B     & 44 & 52800 & 0.74-1.2 & 0.90 & 32.60 & 27.94 & 26.88 \\
V     & 26 & 23400 & 0.72-1.2 & 0.83 & 32.55 & 27.29 & 26.50 \\
r     & 26 & 23400 & 0.72-1.2 & 0.83 & 31.96 & 26.58 & 25.85 \\
I     & 18 & 16200 & 0.60-1.0 & 0.70 & 31.81 & 25.85 & 25.27 \\
\hline
\end{tabular}

(1) : Magnitude limits and zero-points are provided in 
natural magnitude systems  (see text)\\
(2) : Magnitude limits are computed inside apertures with 
diameters of 2$\times$ FWHM without aperture correction 
 (see section 2.3.1).
\end{table*}
\subsection{Data Reduction}
\subsubsection{Flat-fielding and coaddition}
 The single raw frames have  been bias-subtracted and flat-fielded. To correct
 the flat-field pattern, a ``super-flat-field'' was obtained 
 by using all the dithered images and computing an illumination map 
 with a median filtering at 2.5$\sigma$. \\
 The cosmic rays identification was carried out using the cosmic filtering 
 procedure of MIDAS. A pixel is assumed to be affected by a cosmic ray if its
 value exceeds 4$\sigma$ of the mean flux computed for the 8 neighboring
 pixels. These automatic identifications, the positions of known
 detector defects  and of additional events like
 satellites
 trails and elongated radiation events detected by visual inspection,
 were used to build   a ``mask'' frame with ``0'' values for these
 pixels and ``1'' for the others.
 This defect mask is multiplied by the normalized flat-field to produce 
 the weight map of the individual frames. 

 To perform the coaddition of all frames, we have applied the 
 same algorithm  used for the combination of the
  HDF frames (Williams et al., 1996),  known as ``drizzling'' (Fruchter \& Hook, 1998).
  The sky-subtracted, dithered images are flux scaled, shifted and rotated
  with respect to a reference frame corrected 
 for atmospheric extinction.
  Given the good sampling of the instrument PSF by the CCD pixels, we preserved
  the input pixel size in the output image.
  The intensity in one output pixel, $I(x,y)$, is defined as 
\begin{equation}
 I(x,y) = \frac{\Sigma f_{xy}i(x,y)w_{xy}}{\Sigma f_{xy}w_{xy}}
\end{equation}
 where $i(x,y)$ is the intensity in the input pixels, $f_{xy}$ is the overlapping 
 area for the input and output pixels and $w_{xy}$ is an input weight 
 taken into account the flux-scaling, r.m.s. and also
 the weight map  of a single frame. 

 The different number of dithered frames contributing to each  
 pixel of the coadded image produces a non homogeneous noise through 
 the "drizzled" frame. 
 Each combined frame is accompanied by a combined weight map, 
 produced during the  drizzling analysis, which represents the
  expected inverse variance  at each pixel position. 
 
\subsubsection{Photometric Calibration}
The photometric calibration was obtained from several standard stars
from Landolt (1992) observed in the same nights and at similar airmasses.
For each standard field, total magnitudes were computed by using a
fixed aperture (2.5 arcsec)  corrected at 6 arcsec by using the star
with the more stable photometric curve of growth.

The $B'V'R'I'$ magnitudes are calibrated in the ``natural'' system
defined by our instrumental passbands. The zero points of our
instrumental system were adjusted to give the same $BVRI$ magnitudes
as in the standard JKC system for stellar objects with $B-V=V-R=R-I=0$.
The colour transformations used for stellar objects are : 
$B'= B -0.14 \times (B-V)$, $V'= V  -0.08 \times (B-V)$,
$R' = R  -0.33 \times (V-R)$ and $I' = I + 0.01 \times (V-I)$. 

In order to derive the zero-point fluxes, for each scientific frame 
the closest (in time) standard field has been analyzed and reduced to the
same airmass.  The flux-scale  factor corresponding to this scientific
frame is used to provide the final-zero-point.   
The zero-point estimations for all standard fields are consistent within
0.03 mag and are given in Table~\ref{tobs}. The following extinction
coefficients were adopted:
$k_B = 0.22, k_V = 0.11, k_r = 0.05, k_I = 0.02$. \\
Finally, the transformations to the AB systems are given by these relations :
$I_{AB}=I'+0.47$,  $r_{AB}=R'+0.21$, $V_{AB}=V'$, $B_{AB}=B'-0.16$.

\subsection{Data extraction}
\subsubsection{Object Detection and Magnitudes}    
The analysis of ``drizzled'' images were performed with the SExtractor
image analysis package (Bertin \& Arnouts, 1996).

The detection and deblending of the objects has been carried out
using as a reference the weighted sum of the $B~V~r~I$ bands
(in the following {\it combined image}).
The weight assigned to each band is proportional to the signal-to-noise of
the sky in the individual weight-map.

The use of the combined image insures an optimal detection of normal or 
high redshift objects but also of any peculiar objects with strong emission
lines in the bluest bands.
 The object detection is performed by convolving the combined image with the
PSF and thresholding at 1.1$\sigma$ of the resulting background RMS.
The catalogue of the detections is then applied to each of the four individual
frames.

To estimate the total magnitudes, we used the procedure  described by
Djorgovski et al. (1995). For the brighter objects, where the isophotal
area is larger than 2.2 arcsec aperture (17 pixels)
(corresponding roughly to $B \le$25, $V \le$24.5, $r \le 24$ and
 $I \le 23.25$), we use an isophotal magnitude above a surface brightness
 threshold of 1.$\sigma$ of the sky noise. For objects with smaller isophotal 
 area, we  used  the 2.2 arcsec aperture magnitude (corresponding to 2-3 FWHM),
 corrected to a 5 arcsec aperture by assuming 
 that the wings follow a stellar profile. The corrections are
 independent of the magnitude and correspond to -0.15, -0.14, -0.13, -0.09 mag
  for B, V, r, I respectively. The validity of this assumption has already
  been  tested in previous similar deep images (Smail et al. 1995).  
Finally, the magnitudes were corrected for galactic absorption
with $E(B-V)=0.03$  by using $A_B / A_V = 1.33$, $A_r / A_V = 0.83$
and  $A_I / A_V = 0.59$ where $A_V = 3.1 \times E(B-V)$. 

 In this procedure several objects detected in the combined image are
  either too noisy to obtain  a reliable magnitude or undetected in 
  the individual drizzled frames.
 Objects with a computed magnitude within a 2.2 arcsec aperture fainter
 than the expected magnitude limit at 2$\sigma$   
 (i.e. $B\sim 27.57$, $V\sim 26.95$, $r\sim 26.29$, $I\sim 25.57$)
have been assigned an upper-limit magnitude. 

\subsubsection{Star-galaxy Separation} 
 For the star-galaxy separation, we used the classifier provided 
 by SExtractor  applied to the I band, to which corresponds the best PSF.
 This classifier gives output values in the range 0-1 (0 for galaxies 
 and 1 for stars). 
 For I$\le$24 an object is defined stellar if the classifier has a value 
 larger than 0.9. For fainter magnitudes no star/galaxy separation is 
carried out. 
 The combination of bright cut-off in magnitude and high value for
 stellar index insures that the stellar sample is not contaminated by
 unresolved galaxies. In section 4.3, the efficiency of our criterion
 is compared with HST data. \\
%
\begin{figure*}
\resizebox{12cm}{!}{\includegraphics{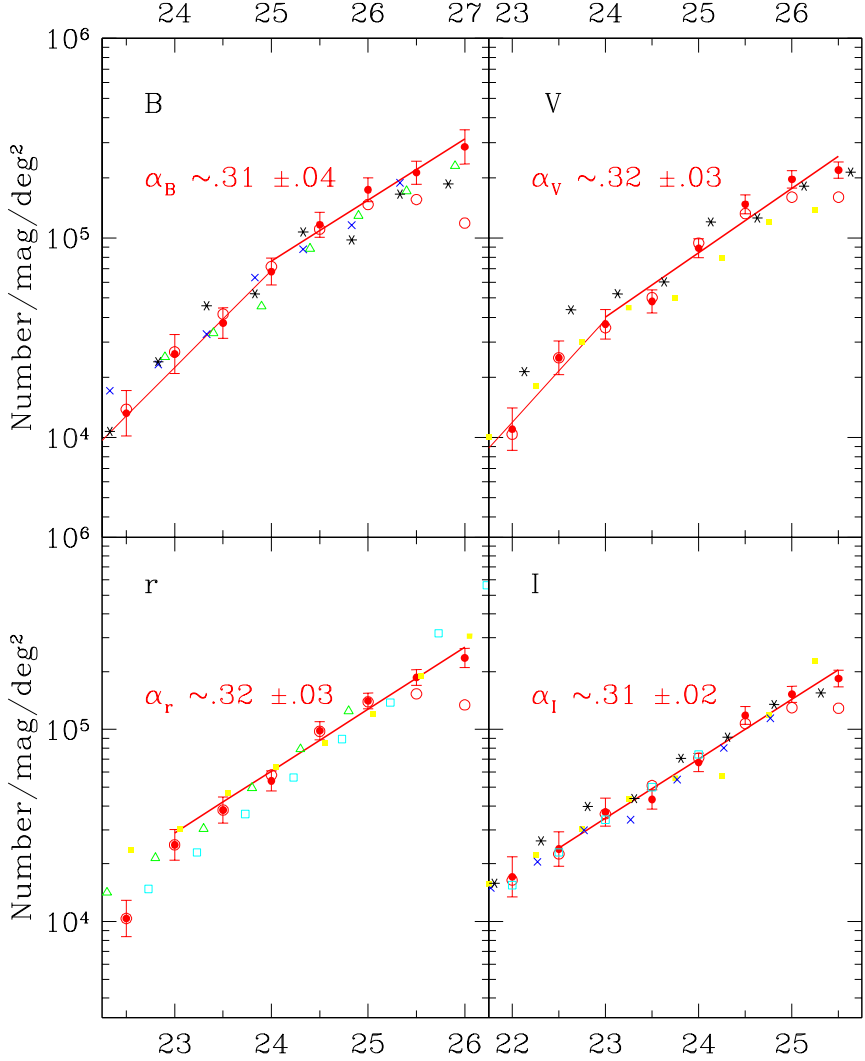}}
\hfill
\parbox[b]{55mm}{
\caption{Differential galaxy number counts for the B, V, r and I bands. 
Open circles show the raw counts and filled circles show the corrected 
counts for incompleteness. The error bars include the correction factor, 
the density fluctuation and poisson errors.
The large solid lines give the best fits for the slope. The previous works 
are drawn with different symbols. They include the data from 
Tyson (1988, open squares), Lilly et al. (1991, crosses), 
 Metcalfe et al. 
(1995, triangles), Smail et al. (1995, filled squares), 
Williams et al. (1996,  stars). Our $r$ filter is significantly different
 to the R Cousins, thus we have applied to the previous works a shift of 0.2 
 ($r = R - 0.33 (V-R)$, by assuming $<V-R> \sim 0.6$).  
\label{fcounts}
        }
                }
\end{figure*}
\begin{table*}	
\caption{Number counts and errors } \label{terr}
\begin{tabular}{ccccccc}
\hline
magnitude &Number counts $(N_{cor})$&$\sigma_{tot}$&Completeness&$\sigma_{pois}/N_{cor}$&$\sigma_{clust}/N_{cor}$&$\sigma_{comp}/N_{cor}$ \\
     &$(10^{4}\ deg^{-2}.mag^{-1})$ & $(10^4)$& factor & (\%)            & (\%)             & (\%)  \\
\hline
 B   &       &      &         &      &      &      \\
\hline
23.5 & 1.32  & 0.35 & 0.96    & 0.9  & 26.2 & 1.2  \\
24.  & 2.62  & 0.59 & 0.98    & 0.6  & 22.5 & 0.9  \\
24.5 & 3.75  & 0.66 & 0.91    & 0.5  & 17.5 & 0.7  \\
25   & 6.78  & 1.05 & 0.95    & 0.4  & 15.4 & 0.5  \\
25.5 & 11.65 & 1.69 & 1.05    & 0.3  & 14.5 & 0.4 \\
26   & 17.42 & 2.39 & 1.18    & 0.3  & 13.7 & 0.4 \\
26.5 & 21.25 & 2.82 & 1.36    & 0.3  & 13.3 & 0.4 \\
27   & 28.58 & 5.64 & 2.41    & 0.3  & 19.7 & 0.5 \\
\hline
 V   &       &      &         &      &      &      \\
\hline
23.  & 1.10  & 0.27 & 1.06    & 1.0  & 24.4 & 1.4  \\
23.5 & 2.51  & 0.49 & 1.00    & 0.6  & 19.4 & 0.9  \\
24.  & 3.69  & 0.63 & 1.04    & 0.5  & 17.0 & 0.8  \\
24.5 & 4.81  & 0.64 & 0.96    & 0.4  & 13.2 & 0.6  \\
25   & 8.89  & 0.97 & 0.95    & 0.3  & 10.9 & 0.5  \\
25.5 & 14.75 & 1.60 & 1.11    & 0.3  & 10.9 & 0.4  \\
26   & 19.67 & 1.98 & 1.23    & 0.2  & 10.1 & 0.3  \\
26.5 & 21.88 & 2.06 & 1.37    & 0.2  & 9.4  & 0.3  \\
\hline
 r   &       &      &         &      &      &     \\
\hline
22.5 &  1.04 & 0.23 & 1.00    & 1.0  & 21.8 & 1.3 \\
23.  &  2.51 & 0.46 & 1.00    & 0.6  & 18.3 & 0.9 \\ 
23.5 &  3.81 & 0.59 & 1.00    & 0.5  & 15.4 & 0.7 \\
24.  &  5.40 & 0.65 & 0.93    & 0.4  & 12.1 & 0.6 \\
24.5 &  9.85 & 1.08 & 1.01    & 0.3  & 10.9 & 0.4 \\
25   & 14.09 & 1.31 & 1.01    & 0.3  & 9.3  & 0.4 \\
25.5 & 18.64 & 1.76 & 1.22    & 0.3  & 9.4  & 0.4 \\
26   & 23.55 & 2.69 & 1.75    & 0.3  & 11.4 & 0.4 \\
\hline
 I   &       &      &       &      &      &      \\
\hline
22.  & 1.71  & 0.41 & 1.04  & 0.8  & 24.0 & 1.1  \\ 
22.5 & 2.39  & 0.49 & 1.06  & 0.7  & 20.6 & 0.9  \\
23.  & 3.71  & 0.62 & 1.02  & 0.5  & 16.7 & 0.8  \\
23.5 & 4.32  & 0.50 & 0.90  & 0.4  & 11.6 & 0.6  \\
24.  & 6.73  & 0.74 & 0.95  & 0.4  & 11.0 & 0.5   \\
24.5 & 11.82 & 1.27 & 1.10  & 0.3  & 10.7 & 0.4  \\
25   & 15.24 & 1.47 & 1.17  & 0.3  &  9.6 & 0.4  \\
25.5 & 18.41 & 1.81 & 1.43  & 0.3  &  9.8 & 0.4  \\
\hline
\end{tabular}
\end{table*}

\subsubsection{Astrometry}
The astrometry uses a tangent-projection to transpose 
the pixel coordinates to RA and Dec (equinox 2000).
The transformation has been calibrated with 15 objects, well distributed
over the whole SUSI field, selected in the
APM catalogue  at the following URL: \\
{\tt http://www.ast.cam.ac.uk/$\sim$apmcat}.\\

A first solution of the astrometry was computed with a polynomial fit
which provides a residual error lower than 0.1 arcsec. Then the 
astrometric parameters have been derived for a tangent-projection
solution and are again consistent with APM at better than 0.1 arcsec.   
To check the reliability of our astrometry, we have compared with 
a WFPC2 image close to the field. In the overlapping region, 108 common 
objects are detected. 
A systematic shift of 0.69 $\pm$ 0.10 arcsec in alpha and -0.70 $\pm$ 0.10 
in delta is observed. The relative astrometric error  is close to 0.1
arcsec and  the systematic  uncertainty is compatible with the value given by the
APM catalogue  ($\sim$ 0.5 arcsec).

\subsection{The Catalogue}     
All the data used in this paper are available in a catalogue present at 
the following URL address: \\
{\tt http://eso.hq.org/research/sci-prog/ndf}. \\
The catalogue is accompanied by a description file.
The identification number, position (pixels and astrometric coordinates), 
total magnitudes and errors in each band are given. Star/galaxy classification and
morphological parameters have been obtained from the I band. 
To estimate colors the I frame has been smoothed to match the seeing 
of the other bands (see section 3.2).
   
\begin{figure*}
\resizebox{0.45\hsize}{!}{\includegraphics{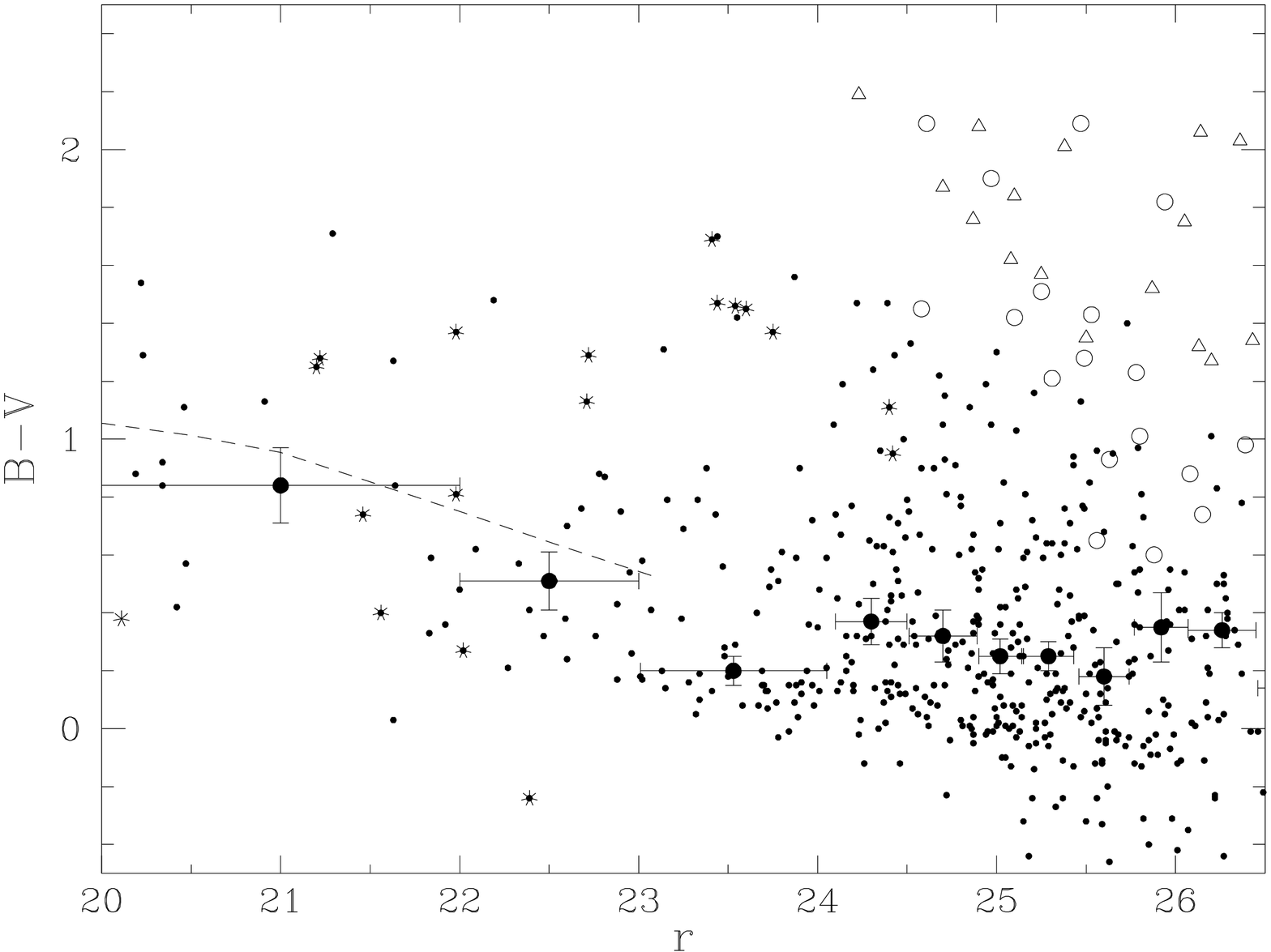}}
\resizebox{0.45\hsize}{!}{\includegraphics{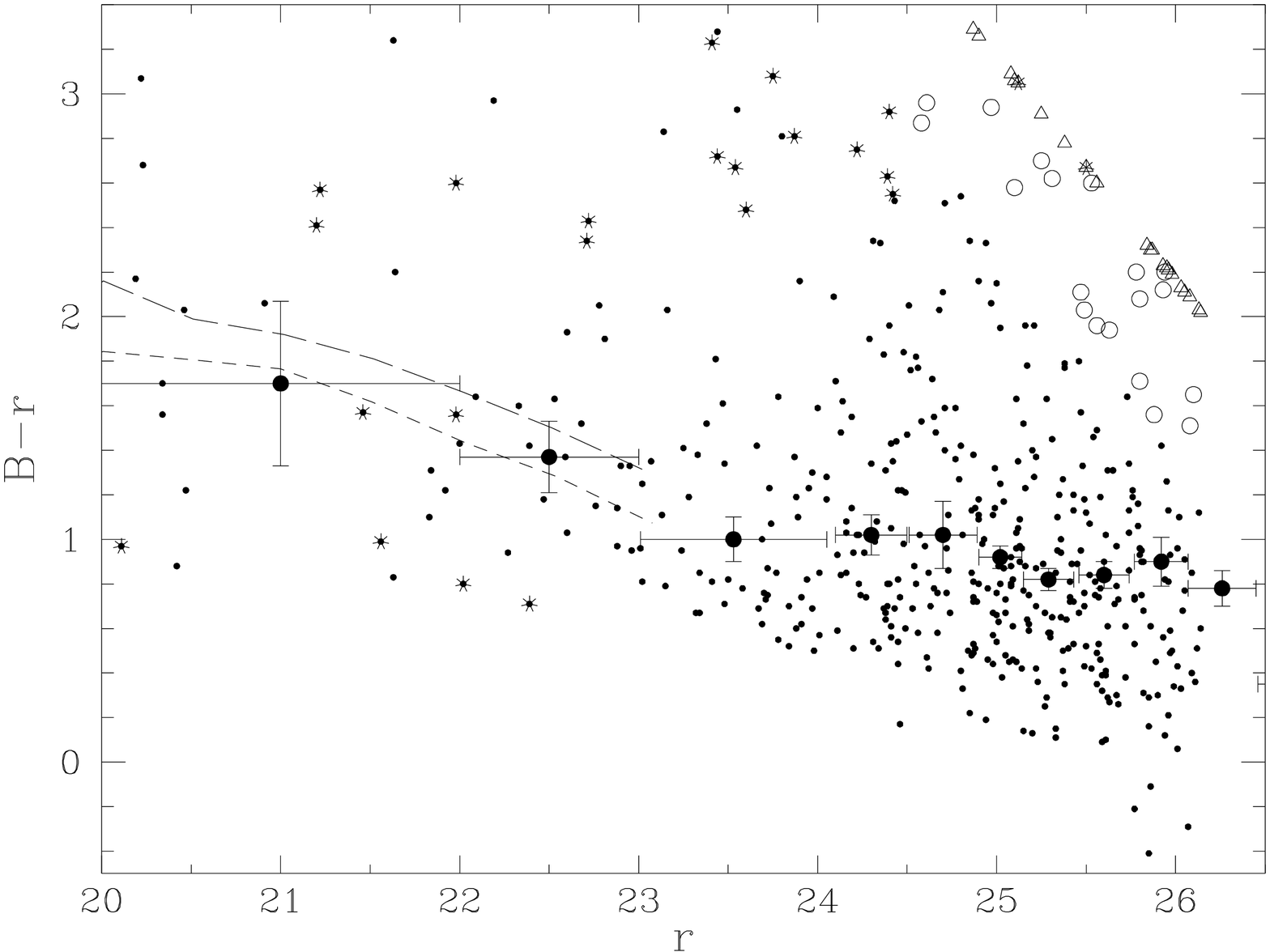}}
\caption{Color-magnitude diagram for $(B-V)$ versus $r$ and $(B-r)$
versus $r$ in our natural systems. The colors are computed in fixed aperture of 15 pixels. 
Open circles represent the colours at 2$\sigma$, open triangles the 
colours at 1$\sigma$ and stars show the stellar objects. 
The large filled points are the medians colors. The horizontal bars show 
the extent of magnitude bin. For magnitudes fainter than 23, the medians
are computed for 60 objects per bin. The vertical error-bars give the 
1$\sigma$ error calculated using boot-strap resampling of data.
Comparison with Arnouts et al. (1997,short dashed line) and Metcalfe et al. 
(1995, long dashed line) are shown by applying the color correction given in sect.~2.2.2.
\label{bcol}}
\end{figure*}
\begin{figure*}
\resizebox{0.45\hsize}{!}{\includegraphics{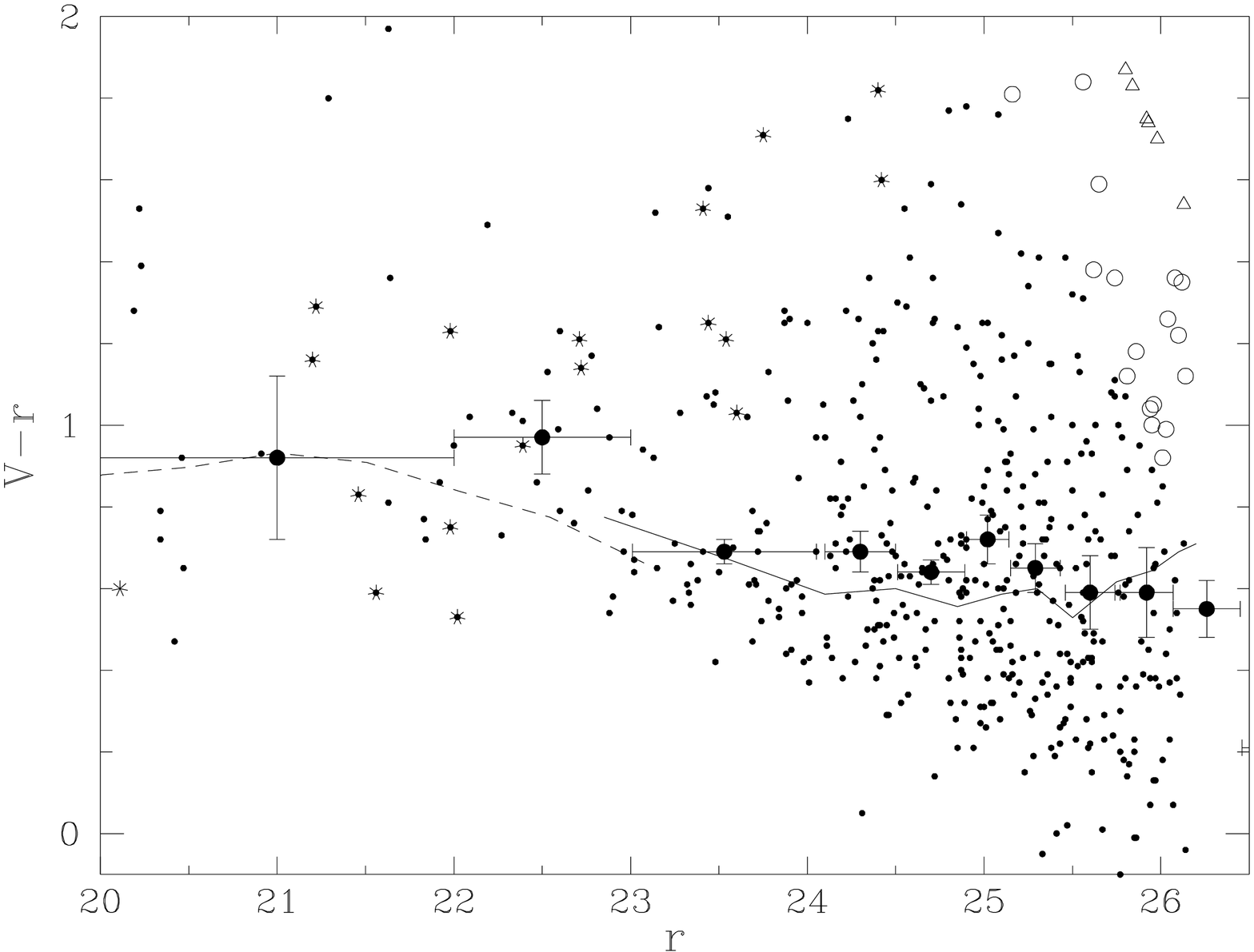}}
\resizebox{0.45\hsize}{!}{\includegraphics{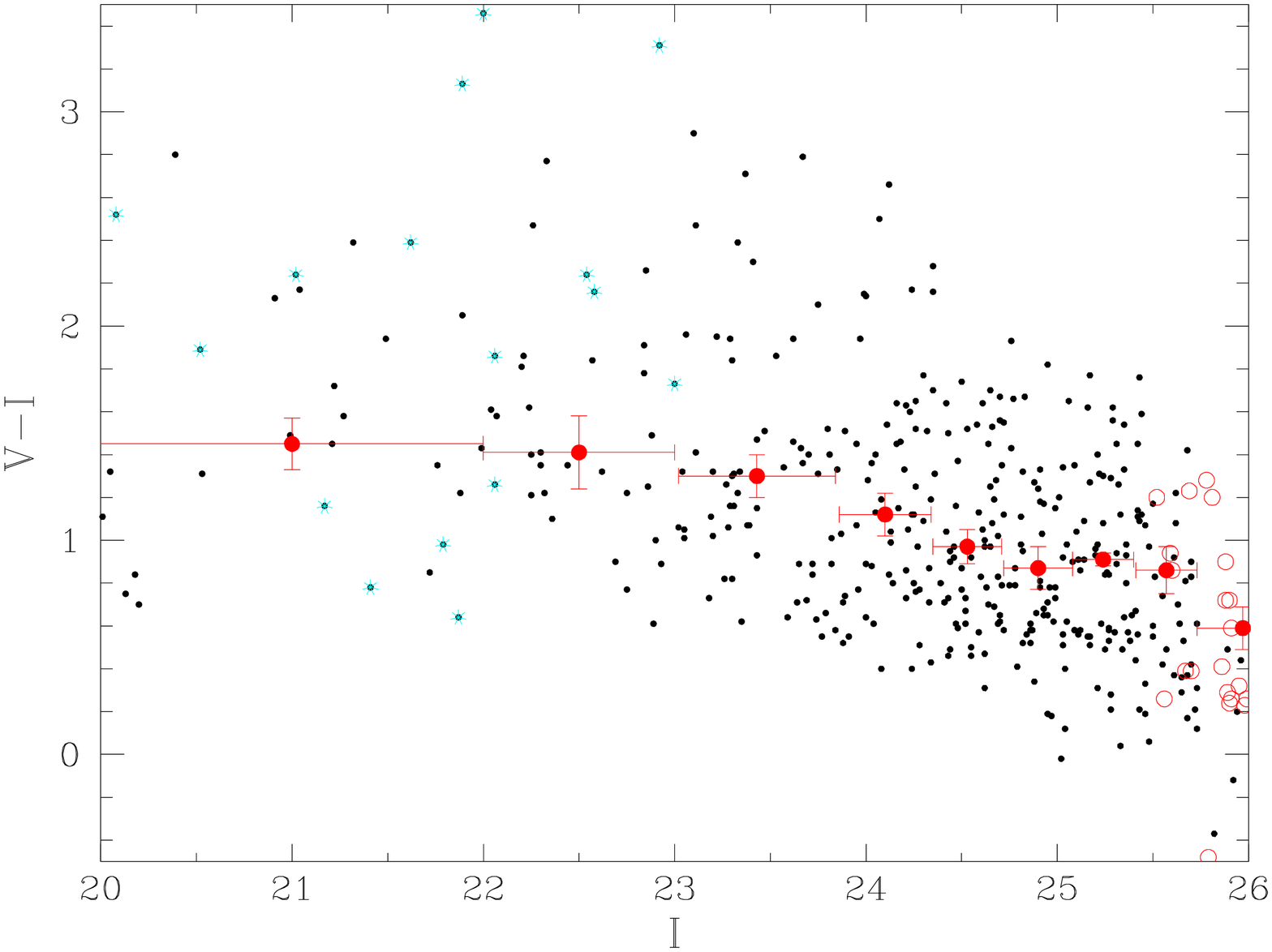}}
\caption{Same as figure~\ref{bcol} for $(V-r)$ versus $r$ and $(V-I)$
versus $I$. Comparison with Arnouts et al. (1997, dashed line) and Smail 
et al. (1995, solid line) are shown by applying the color correction given in sect.~2.2.2.
\label{vcol}}
\end{figure*}
\begin{figure*}
\resizebox{0.45\hsize}{!}{\includegraphics{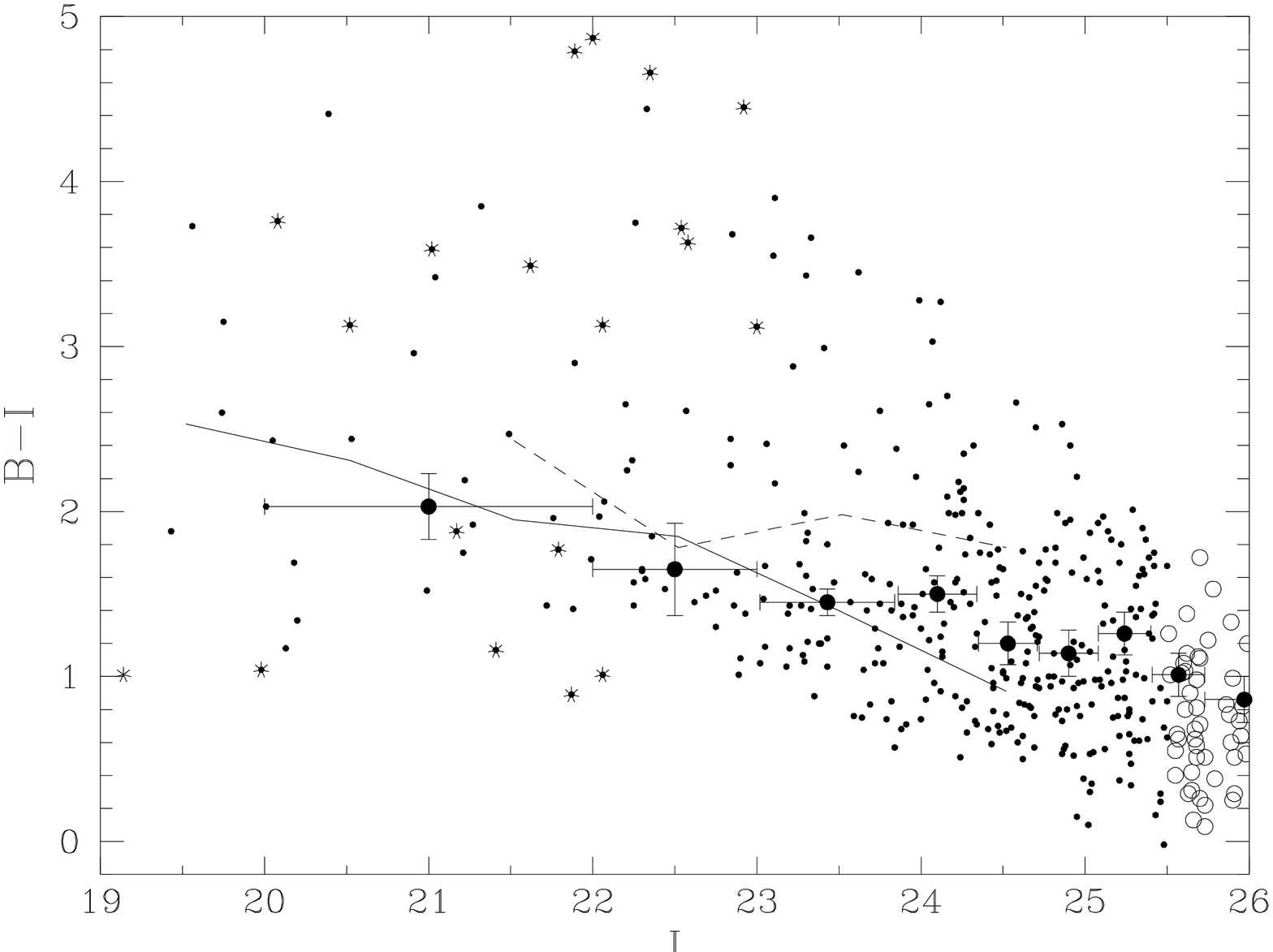}}
\resizebox{0.45\hsize}{!}{\includegraphics{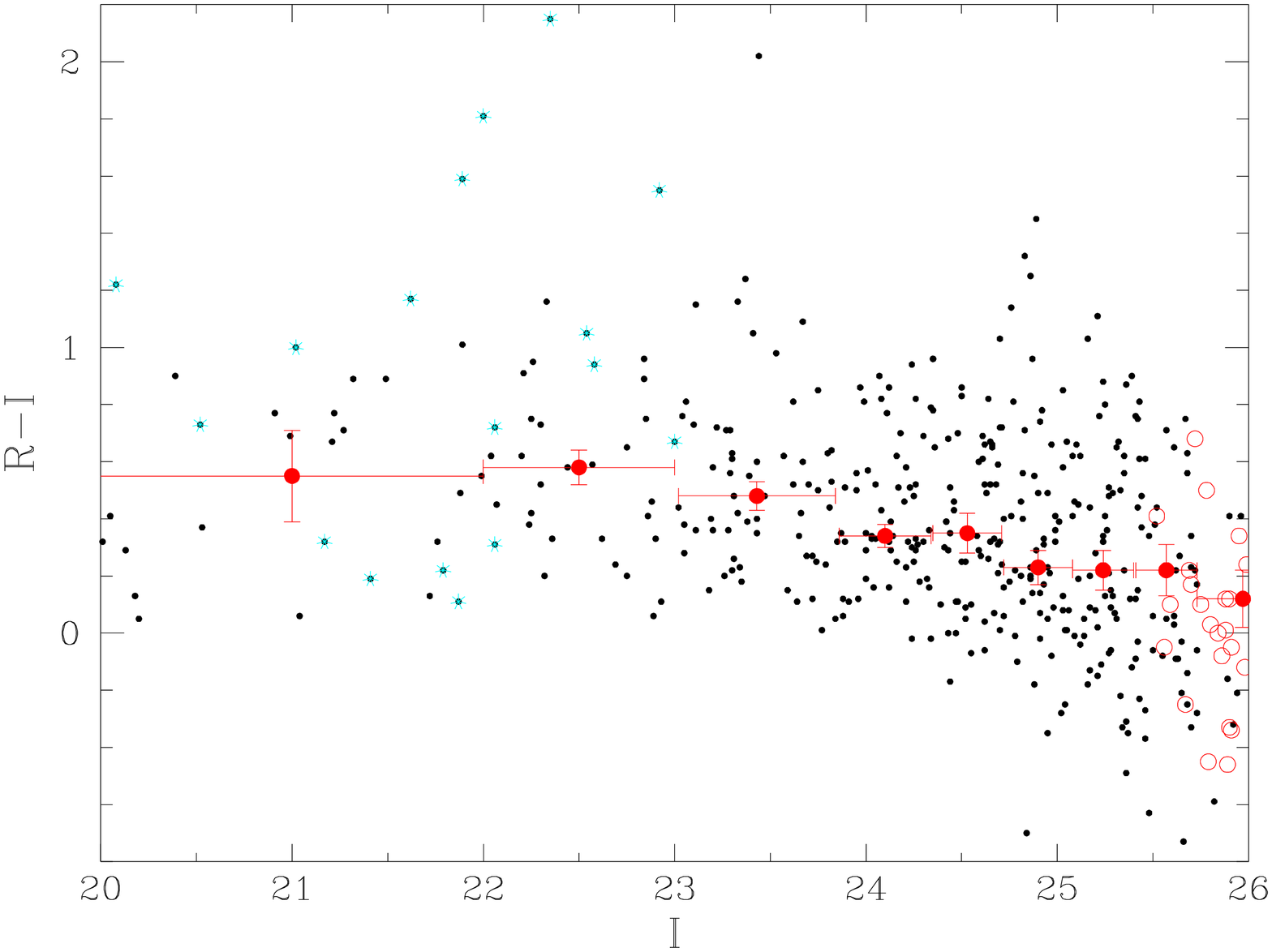}}
\caption{ Same as figure~\ref{bcol} for $(B-I)$ versus $I$, and  $(r-I)$ versus $I$.
Comparisons are shown for the data from  Tyson (1988, solid line) and Lilly et al.
(1991, dashed line) by applying the color correction given in sect.~2.2.2.
\label{rcol}}
\end{figure*}

\section{Counts and colors}
\subsection{Differential Galaxy Counts}
%
%
 Before discussing the galaxy counts, we have to apply a correction
 for completeness at faint magnitudes.  A standard way to estimate the
 fraction of lost sources, crowding, spurious objects, magnitude
 errors, etc ..., is to generate simulations by assuming a typical
 profile for all kinds of galaxies and different cosmological models
 (Metcalfe et al, 1995).  In the present case, we have to take into account
 that the detection of sources has been done by using the
 combined image of the four bands and that the noise in our
 drizzled frame is correlated due to the drizzling algorithm. To
 provide an estimation of the correction  as close as possible to the
 characteristic of our raw data, we have adopted the following empirical 
approach:
\begin{enumerate}
\item{
we extract objects in the magnitude ranges 
23$\le B, V \le $ 25.5 and 23$\le r, I \le$25.0
(by using the ``check-images'' tool of SExtractor software).  At these
magnitudes, the signal-to-noise is greater than 8.
}  
\item {
the object's fluxes are dimmed by factors 4, 6.5 and 10 and added randomly to the
original frames (combined image and $B~V~r~I$ images).  The noise in the 
resulting combined and 
$B~V~r~I$ images is identical to the original data.
} 
\item {
 The algorithm of
 detection is applied to the combined image and the measure of the magnitudes
 is carried out on the $B~V~r~I$ frames. The correction
 factor for each magnitude bin is derived from the fraction of the
 ``simulated'' sources that are detected.}
\end{enumerate}

 The magnitude range of the original sources used in this simulation
has been chosen in a way that a sufficient number of objects between
$24.5-27.5$ is generated, while avoiding a too large crowding at
intermediate magnitudes.  Besides, since the original galaxies have a
mean expected redshift of 0.6-1.0 and we do not expect drastic
differences of colors with respect to the fainter galaxy population
(except for the ellipticals in the B band), this allows us to
ignore any additional k-correction term.  Note that this analysis
doesn't include the Eddington bias and can overestimate the correction
factor.
 The completeness factor ($f = \frac{N_{simul}}{N_{detect}}$) is reported in table ~\ref{terr}.

  In Fig.~\ref{fcounts}, we plot the raw and corrected differential
  galaxy counts as a function of the  magnitude ($N_{cor} = N_{raw} . f$) and we report the 
 corrected number counts in table ~\ref{terr}.
  The error bars include  the quadratic sum of the Poisson noise ($\sigma_{pois}$),
  the uncertainty in the completeness  factor ($\sigma_{comp}$)  
  and the clustering fluctuation due to the small size of the field as
  follows :
\begin{equation}
 \sigma^2_{clust} = <N_{cor}>^2 \frac{A_{\omega}}{\Omega^2} \int \theta_{12}^{-0.8} d\Omega_1 d\Omega_2  
\end{equation}    
\begin{equation}
 \sigma^2_{clust} = <N_{cor}>^2  \ A_{\omega} \ IC
\end{equation}
 where 
 IC is the  integral constraint  computed by a Monte-Carlo method for the
 appropriated size of the NTTDF  field ($IC = 33.58$).
  The relationship between the amplitude $A_{\omega}$ and the 
  magnitude is assumed to evolve as $A_{\omega}(at \ 1^{\circ}) = -0.3 * mag + C_{tt}$,
  where $C_{tt}= 4.4, 4.1, 3.9,  3.8$ for the B, V, r and I bands respectively.
  The individual and  total errors are reported in table ~\ref{terr}. 

  The slopes  are estimated by linear regression. 
 For the B and V bands, the counts show a significant flattening at
 magnitudes fainter than  
 B$\sim$ 25 and V$\sim 24$ with slope values of $\alpha_B = 0.31 \pm 0.04$
  and $\alpha_V = 0.32 \pm 0.03$. 
 The evidence of  breaks in the B and V bands have been  already produced 
 by previous works (Metcalfe et al. (1995), Smail et al. (1995)). 
The low density of bright galaxies in the NTTDF produces an artificially
 steep slope at bright magnitudes, making more apparent 
the breaks in the counts at 25 and 24 in B and V.
 
 The slopes at faint magnitudes in the B and V bands are similar to 
 those  observed  in the whole range in r and I bands : \\
    $\alpha_r = 0.32 \pm 0.03$ with $23 \le r \le 26$ 
  and $\alpha_I = 0.31 \pm 0.02$ with $22.5 \le I \le 25.5$.

  In  Fig.~\ref{fcounts}, our results are compared with the  previous works
  of Tyson (1988), Lilly et al. (1991),
  Steidel et al. (1993), Metcalfe et al. (1995), Smail et al. (1995) and
 Williams et al. (1996).
 At bright magnitudes, the discrepancy is large ($\pm 0.1$ in the
  slope) due to the bias mentioned above. At faint magnitudes
  the slopes  are consistent among the different works. 
\subsection{ Colours of the Galaxies in the Sample}
%
 To measure colours, we have smoothed the I frame to the same effective 
 seeing of the worst frame (B frame) and carried out photometry in 2
 arcsec diameter apertures  with the same center as defined in the 
 summed frame. The colors versus magnitude plots are shown in Fig.~\ref{bcol},
 Fig.~\ref{vcol},Fig.~\ref{rcol}. The 2$\sigma$ and 1$\sigma$ color limits 
 within the 2 arcsec apertures are represented by open circles and triangles 
 respectively. The median include the colour limits 
 at 1 and 2 $\sigma$ and has been computed in bins with 60 objects except
 for bright magnitudes where bins extending from 20 to 22 and from 22 to 23 
 have been imposed. The error bars on the median colours have been computed 
 by using a boot-strap resampling method. A comparison with previous works 
 is shown with solid and dashed lines. This comparison has to be taken with
 caution because the colors are estimated with constant number of objects
 in each bin  and not with a fixed bin.  
 All the color distributions show up to r$\simeq$24-24.5 a blueing trend 
 at bright magnitudes.  
 At deeper magnitudes, the median $B-V$, $B-r$ and $V-r$ versus $r$
(Fig.~\ref{bcol}, ~\ref{vcol}) show stable values. 
 The $B-I$, $V-I$ and $r-I$ versus $I$ (Fig.~\ref{vcol} and Fig.~\ref{rcol})
 show again a blueing trend  but no color stabilization is observed at the
 fainter magnitudes.
 This flattening in the various colors is in good agreement with the 
 observed convergence of the slopes ($\sim$ 0.32) in the differential counts 
 for all the four filters.  \\ 
\subsection{Magnitudes and Colours of the Stellar Objects }
A by-product of this catalogue is a sample of stellar objects. 
As in the case of the HDF, the study of the stellar population
present in the NTTDF can give useful insight in the galactic
structure and/or give constraints on the baryonic dark matter. 
With the NTTDF it is possible to perform a direct comparison of its 
stellar content with the one of the HDF.
The two fields have essentially the same galactic latitude (for
NTTDF $b=53.4$ and for HDF $b=54.8$), but different galactic longitude,
with the NTTDF pointing toward the galactic center ($l=283.6$), 
while the HDF is pointing toward the galactic anti-center ($l=125.6$).
Moreover, they cover essentially the same area in the sky ($\simeq0.0016$ 
square degrees), allowing us to make a direct comparison of the two 
stellar samples.

\begin{figure*}
\resizebox{\hsize}{!}{\includegraphics{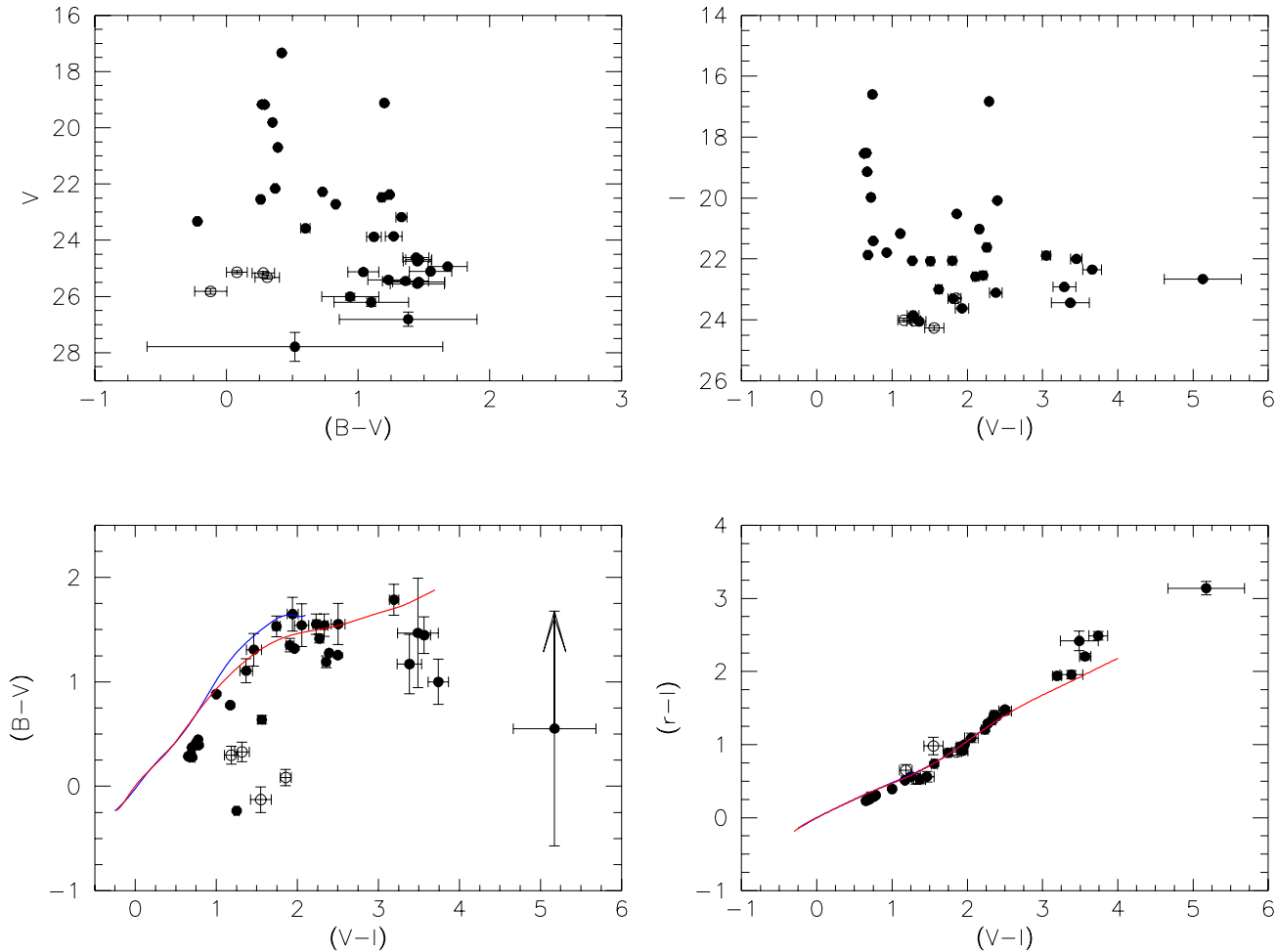}}
\caption{In these plots we present the color-magnitude and color-color
diagrams of the objects classified as stellar in the NTTDF. 
{\it Upper Left.} V {\it vs.}(B-V) color-magnitude diagram. 
{\it Upper Right.} I {\it vs.}(V-I) color-magnitude diagram. 
{\it Lower Left.} (B-V) {\it vs.} (V-I) color-color diagram.
{\it Lower Right.} (r-I) {\it vs.} (V-I) color-color diagram.
Continuous lines are the colors expected for giants/dwarfs 
stars (from Caldwell et al. 1993).}
\label{starcmds}
\end{figure*}

In the upper panels of Fig.\ref{starcmds} we show the V {\it vs.}(B-V) 
and I {\it vs.}(V-I)  color-magnitude diagrams of the objects classified 
as stellar (star/galaxy class$\geq0.9$) in our final catalog.
The objects selected have colors in good agreement with the 
color expected for stellar objects. In the lower panels of 
Fig.~\ref{starcmds} we have put the (B-V) {\it vs.}(V-I) and 
the (r-I) {\it vs.}(V-I) color-color diagrams. 
It can be clearly seen that, especially in the (r-I) {\it vs.}(V-I) 
(where the observational errors are lower), the objects follow very 
closely the colors expected for giants/dwarfs stars (the continuous 
lines are taken from Caldwell et al. 1993), assuring us that the 
classification parameter gave reliable results. 
In all the plots we have also included 4 objects with star/galaxy 
class$\geq0.8$ and $<0.9$ (shown as open dots).
These blue objects have been included since they resemble the 
population of blue objects found by Mendez et al. (1996) in their 
analysis of the stellar content of the HDF: they could be
interpreted as partially unresolved blue galaxies. 
In this sample the object in the (B-V) {\it vs.}V diagram at V$\simeq23.2$ 
is the brightest and bluest member of this sample. 

Of some interest for a follow up spectroscopic observation 
is the object \#606 of the catalogue, having a (V-I)$\simeq5.2$.
The colors of this object, including an infrared (I-K)=3.2 
(Fontana et al. 1998, to be submitted), are consistent with 
those expected for bright low mass stars (Delfosse et al 1998).
This object has also colors very similar to KELU-1, a field 
brown dwarf with a mass below $0.075M/M_\odot$, and a distance from the
Sun of $\sim12$pc (Ruiz, Legget, \& Allard 1997).
Other likely faint brown dwarfs are the 5 objects with $3.2<(V-I)<4.0$;
their catalogue number is: \#93, 112, 249, 579, 600.

In our (V-I) {\it vs.}I diagram we don't have stars below I$\simeq24$, 
even if our detection limit is I$\sim25.6$. This difference could be
explained by the fact that are needed at least 5 times more photons 
to classify objects rather than just detect them (Flynn et al. 1996). 
In the magnitude range $20<$I$<24$ we found 26 objects, with 6 
objects at (V-I)$>3$ ($\sim20\%$ of this sample).  The number of 
objects are approximately 3 times the stars detected in the same 
magnitude range in the HDF (see Flynn et al. 1996, Mendez et al. 
1996, Reid et al. 1996 for different results on the HDF). 
This difference in number may be explained by the 
fact that the NTTDF is pointing in the direction of the galactic
center. It is compatible with the prediction obtained with the galactic model
by Robin and Creze (1986), as available at
{\tt www.ons-besancon.fr/www/modele\_ang.html}.

Finally, The object \#570, classified as stellar-like on 
the basis of the SExtractor algorithm, is listed in the 
catalog by Giallongo et al 1998 (where it is \#12) with colors 
typical of an high redshift candidate. 
Cowie and Hu (1998, private communication) obtained a spectrum 
 at Keck. Although the S/N is low, an M star 
interpretation of the spectral features appears more likely. 
The colors of the object 
(including the $I-K\simeq$1.5 of Fontana et al. 1998) would be 
marginally consistent with a slightly earlier (K5-K7) classification. 
In Sect. 5 (see in particular Fig. 9) it will be shown that few red 
stars display colors similar to high-z galaxies.

\section{Comparison with the Hubble Space Telescope }
\subsection{HST Observations of the same field}

We compared the NTTDF I-band image photometry with the photometry of 
WFPC2 archival images of the same field.
The WFPC2 data consisted in a set of 8 F814W images totaling 
9200 sec of exposure, with the QSO BR1202-0725 centered on CCD \#3. 
Of the four WFPC2 CCD we used only CCD \#2 which was almost completely 
overlapping ($~90\%$ of the frame) on the North-East corner of the NTTDF 
I-band image.
We carried out a completely independent analysis of the WFPC2
data from the NTTDF data, since we wanted to investigate the
consistency
of the two data sets and in particular the impact of the lower angular
resolution of the ground-based observations on the morphological and 
the star/galaxy classification algorithm of SExtractor.

\begin{figure}
\resizebox{\hsize}{!}{\includegraphics{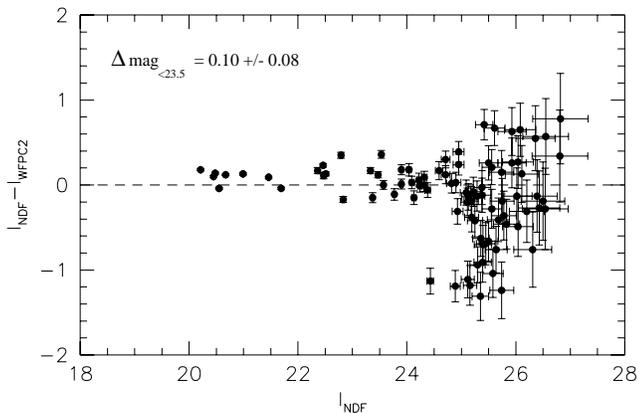}}
\caption{Magnitude comparison of the NTTDF and WFPC2 photometries.
\label{magcomp}}
\end{figure}

The photometry of the objects in the WFPC2 image was performed 
on the coadded and cosmic ray cleaned image. 
We used a detection threshold of $\sim1.0\sigma$ of the sky noise and a FWHM 
of 1.5 pixels, finding 132 objects.
The estimated magnitude limits at $5\sigma$ in an aperture of $2\times$FWHM 
($\simeq0.\!''3$) is I$_{AB}=24.21$.
The data have been calibrated in the STMAG system of the WFPC2, 
correcting the zero point to the AB system with \\
ABMAG=STMAG$-0.819$ (Williams et al 1996).

\begin{figure}
\resizebox{\hsize}{!}{\includegraphics{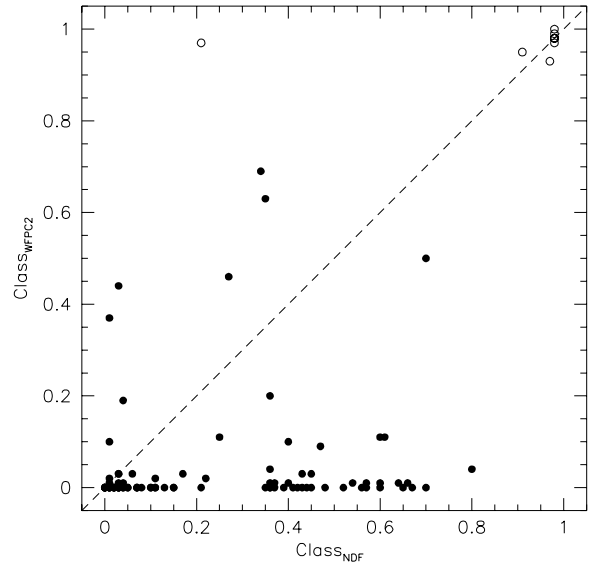}}
\caption{Comparison of the star/galaxy classification in the 
NTTDF and WFPC2 photometries. Open dots are objects classified as
stars in the WFPC2 image, while filled dots are objects classified as 
galaxies in WFPC2.
\label{classcomp}}
\end{figure}
\begin{figure}
\resizebox{\hsize}{!}{\includegraphics{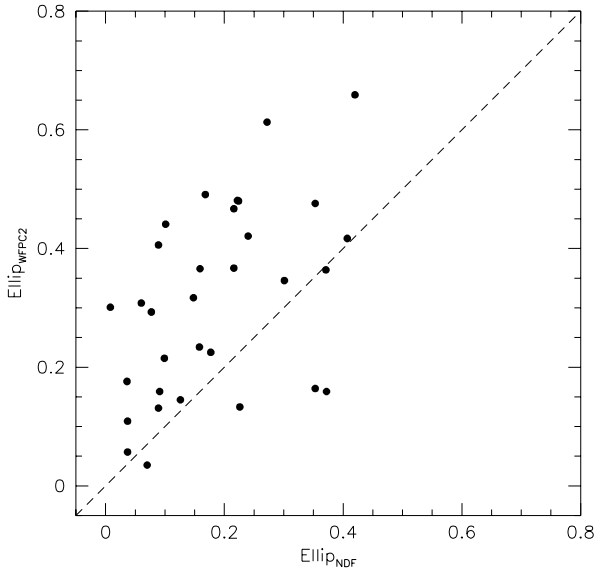}}
\caption{Comparison of the ellipticity estimation in the 
NTTDF and WFPC2 photometries.
\label{shapecomp}}
\end{figure}

\subsection{Matching WFPC2 and NTTDF catalogue of objects}
We used the astrometric solution for the WFPC2 data (obtained with the
{\tt metric} command of IRAF) to match the objects in the NTTDF.
After eliminating a residual shift of  $0.\!''69\pm0.10$ in RA, 
and $-0.\!''70\pm0.10$ in Dec (these values are within the expected 
values for the HST pointing precision), we found 108 common objects 
in the two lists.
The matching has been done finding the nearest object in the WFPC2 catalog 
for each object in the NTTDF catalog, with a maximum tolerance of 1 arcsec. 

In the common area only 10 WFPC2 objects were not found in the NTTDF image,
and 40 NTTDF objects were not recovered in the WFPC2 image.
We verified the nature of these non-matched objects looking the
morphology of each of them in both images.
In the case of the 10 non-matched objects detected in the WFPC2 image we 
found that one of them was a blend of a relatively bright star and a galaxy 
in the NTTDF frame, while 4 other objects were clearly ``blobs'' (star formation 
regions?) associated to some extended objects.
The remaining 5 objects were just at the detection limit in the WFPC2 image
and barely visible objects in the NTTDF image, just below the detection 
threshold.
Using this number, we conclude that a reasonable estimate of the frequency of 
blended objects in the NTTDF field should be $\simeq1\%$.

The large number of objects present in the NTTDF catalogue and not found in
the WFPC2 image (40) is due to the fact that the NTTDF catalog was 
obtained using the summed image of the four different bands.
This means that many of the objects were not present also in the I-band 
image of the NTTDF.
In fact, a direct inspection of the two images at the location of the 40
non-matched objects revealed that all the objects missed in the WFPC2 
image were at the detection limit in the NTTDF, and that no bright object
was missed:  16 out of 40 objects were classified as having upper limit 
magnitudes in the
NTTDF catalog, and the remaining 24 had a magnitude below $\simeq 25.0$.
All these objects are barely visible in the WFPC2 image.

\subsection{Magnitude and morphology comparison}
Fig.~\ref{magcomp} shows the comparison of the photometry of the NTTDF I-band 
image with the WFPC2 in the almost equivalent F814W band. 
We selected only the 86 object having the SExtractor flag $\le3$
(only isolated objects or with marginal blending) in both the NTTDF and WFPC2,
and excluding all the objects with  upper limits.
The isophotal magnitude difference between NTTDF and WFPC2, computed
for objects brighter than 23.5, gives a mean value of $0.10\pm0.08$ mag.
This difference is compatible with the uncertainties in the absolute
calibration of the two photometric systems.

Despite the broadening of the stellar profiles due to the effect of the seeing,
the comparison of the the star/galaxy classification of the NTTDF objects
and the one of the WFPC2 shows a remarkable agreement.
In Fig.~\ref{classcomp} we show the two classifications using filled circles
 for objects with the classifier $<0.90$ (galaxies) in the WFPC2 frame and
 with open circles for objects with $\ge0.90$ (stars) in WFPC2.
The figure shows that of the 9 objects classified as stars down to the 25 mag
in WFPC2, only one is misclassified as a galaxy in the NTTDF.
This means that the residual contamination of stars down to the magnitude 25
 should be $\simeq1.5\%$ (1 star out of 78 galaxies).

%
\begin{figure*}
\resizebox{12cm}{!}{\includegraphics{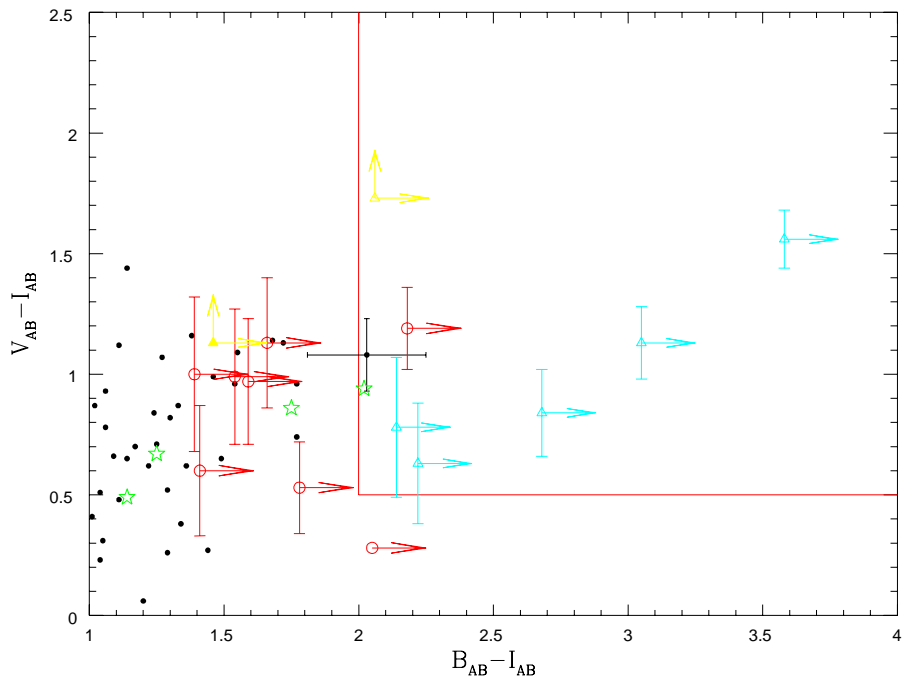}}
\hfill
\parbox[b]{55mm}{
\caption{Color criteria to select star forming galaxies
 at redshift 3.8 $\le z \le $ 4.4. Triangles show objects with $B-I$ upper limits  distant less
 than 1$\sigma$ from the color selection criterion, circles show objects with upper limits distant
 between 1 and 2$\sigma$ from the selection criterion.  
 Both are considered as upper-limits (defined with arrows). The galaxy with 
 B and V upper-limit and filled triangle (\# 596 in the catalogue) is spectroscopically
  confirmed at  $z\simeq 4.2$.  Stars (with stellarity index $\ge$0.9) are shown with star
  symbols. For clarity, error-bars are put only for galaxies matching 
 the color criteria.   
\label{fz4}}}
\end{figure*}

The comparison of morphological shape parameters (minor and major axis,
inclination ($\theta$), ellipticity) of the NTTDF with the one of the
WFPC2 revealed that ground-based observations recover the  shape 
parameters within an accuracy of 30\% only.
We give an example in Fig.~\ref{shapecomp} where we compare the ellipticity
parameter for galaxies brighter than 25 mag and with a total area greater 
than 25 WFPC2 pixels (38 objects). 
We cannot find object with ellipticity above $\sim0.4$ in the NTTDF
while in the WFPC2 we can easily reach a value of $\sim0.75$:
this ``roundization'' effect is first due to the atmospheric seeing
and second to the coadding of dithered frames of different image quality.

\subsection{The Hubble Deep Field}
 The Hubble Deep Field project (Williams et al 1996) has set a new benchmark
for multicolor deep survey. The total integration time dedicated to the HDF
was significantly longer than for the NTTDF (506950 seconds versus 
115800). A measure of the relative speeds in collecting photons 
can be made for the 
B and I bands, which are the only ones to be relatively similar. The
HST F450W band is 
wider and shifted by about 30 nm to the visual than its SUSI 
counterpart, while
the F814W is close to the corresponding SUSI I band. The integration 
times on the HDF are a factor of 2.3 and 7.6 longer in the two 
bands respectively.
The factors which enter in the determination of the limiting magnitudes are,
beside the exposure times, the ratio of collecting areas between the
two telescopes and the efficiency of the light path, including mirrors,
instrument optics, filter pass bands and detector. By using the values given 
by Williams et al. 1996 for the HST (their figure 2) and the computed values
for SUSI, we derive a relative efficiency SUSI/ HST, 
including the ratio of the collecting areas, of 5 in both B and I respectively.
The additional two important factors in determining the limiting magnitude 
are the sky surface brightness
 and the size of the images. For stellar objects, when the 
magnitude is computed within 2 FWHMs of the Gaussian fit, the HST takes
maximum advantage of the superior image quality because the background is both 
of lower surface brightness and computed over a smaller area. The limiting 
magnitudes of the HDF are a factor of 6 and 11 deeper than those of the 
NTTDF. The gain is reduced to a factor of 2 and 7 for diffuse 
objects with 2 x FWHM $\geq$ 1 arcsec. 

\section{The multicolor selection for high redshift galaxies}

Steidel \& Hamilton (1993) used a combination of three filters U, G and
 ${\cal R}$ to select star forming galaxies at 2.8$\le z \le $ 3.4.
 At these redshifts, due to the shift of the Lyman continuum
 break in the UV band,
 the U flux is attenuated and the $U-G$ color is strongly reddened 
 while the  $G-{\cal R}$ color still reflects a flat spectrum. \\
 To select galaxies at higher redshifts, a different set of filters is 
 required. At $z\sim$4, the combination of Lyman $\alpha$ clouds  absorption
 and the Lyman continuum  reduces significantly the observed flux also in the 
 B filter. Giallongo et al. (1998) have shown that an optimized 
 set of filters can be found to detect star forming galaxies at $z\simeq 4$
 and to reduce the confusion with low redshift galaxies of similar colors.
 This set is based on the B,V,r and I bands, and corresponds to the one adopted in the present work.
 An optimized detection  of star forming galaxies in the redshift range 3.8$\le z \le$4.4
 can be obtained by using the following criteria : $(B-I)_{AB}\ge$2, 
 $(V-I)_{AB}\ge$0.5 and $(r-I)_{AB}\le$0.1.
 In Figure~\ref{fz4}, we have  applied the criteria to our catalogue. 
Two approaches have been followed to select galaxy candidates up to $r\le 26$.
\begin{itemize}
\item A: Galaxies with measured colors (or upper limits) 
located inside the area defined by the above criteria (8 galaxies).
\item B: The above sample and galaxies with $(B-I)_{AB}$ upper limits falling less than
 $1 \sigma$ outside the selection criterion (15 galaxies).
\end{itemize}
The photometric redshifts for the galaxies of this field will be
estimated in a forthcoming paper  (Fontana et al., 1998) by comparing
the colors (including J and K) with a  library of synthetic galaxy spectra as
described 
 by Giallongo et al. (1998).
Due to their faint magnitudes,
the spectroscopic  confirmation of our candidates requires 
an 8-m class telescope.
Only one of the galaxies (object \# 596 in our catalog) is 
included in the spectroscopic sample by Hu et al. (1998) and Hu
(1998,private communication) 
and has a spectroscopic redshift of $z\simeq 4.19$. This galaxy has upper
limits in the B and V bands and a magnitude $r \simeq 25.9$. \\
The number of galaxy candidates in different magnitude intervals are
given in Table~\ref{thz1} and Table~\ref{thz2} for both classes $A$ and $B$
described above.  The surface density at $r\le$25, is in the range 
0.7-0.9 arcmin$^{-2}$ (case $A$ and $B$ respectively). At $r\le$26, the density 
 increases to 1.4-2.7 arcmin$^{-2}$.  

%
\begin{table}
\caption{Number distribution of galaxies in redshift range 
$3.8\le z\le4.4$ located inside the color selection criteria 
(case A with 8 objects).
\label{thz1} }
\begin{tabular}{ccccc}
\hline
$r$ &  Num.  & up.-lim. & up.-lim. & cum. dens. \\
    &        &     B    & B \& V   & (arcmin$^2$) \\
\hline
24$\le r \le$ 24.5 & 1 & 0  & 0  &  0.2  \\
24.5$\le r \le$ 25 & 3 & 2  & 0  &  0.7  \\
25$\le r \le$ 25.5 & 2 & 2  & 0  &  1.1  \\
25.5$\le r \le$ 26 & 2 & 2  & 1  &  1.4  \\ 
\hline
\end{tabular}
\end{table}
\begin{table}
\caption{Same as Table~\ref{thz1} for case B (with 15 galaxies).
\label{thz2} }
\begin{tabular}{ccccc}
\hline
$r$ &  Num.  & up.-lim. & up.-lim. & cum. dens. \\
    &        &     B    & B \& V   & (arcmin$^2$) \\
\hline
24$\le r \le$ 24.5 & 1 & 0  & 0  &  0.2  \\
24.5$\le r \le$ 25 & 4 & 3  & 0  &  0.9  \\
25$\le r \le$ 25.5 & 4 & 4  & 0  &  1.6          \\
25.5$\le r \le$ 26 & 6 & 6  & 2  &  2.7          \\ 
\hline
\end{tabular}
\end{table}
%
 Assuming a uniform redshift distribution of the sample in the range 
 3.8$\le z \le $4.4, the comoving galaxy number density at $<z> \simeq 4.1$
 for $r\le25$ is estimated to be $6.5-8.1~ 10^{-4}  h^3 Mpc^{-3}$ (q$_0$=0.5)
 and $1.3-2.4 ~ 10^{-3} h^3 Mpc^{-3}$ (q$_0$=0.5) for r$\le 26$.  \\
  
 The star formation rate history at $z\simeq$ 4.1 can be derived from the
 luminosity in the I band corresponding to the restframe wavelength 1500 \AA.
 We adopt the  conversion factor (from the UV  luminosity at 1500 \AA \ 
 to the SFR)  of Madau et al. (1996). For a Salpeter IMF
 (0.1$\le M \le 125 M_{\odot}$) with constant SFR, solar metallicity and 
 age range of 0.1-1 Gyr, a galaxy with $ SFR = 1 M_{\odot} \cdot yr^{-1}$
 produces   $L_{1500} = 10^{40.15 \pm 0.02} \  erg.s^{-1} \cdot \AA^{-1}$. \\
 The brightest galaxy in our sample  has a magnitude of 
 $I_{AB}=24.5$ corresponding to a  $SFR=15.0 \pm 1.5 \ (55 \pm 8) h^{-2} 
 \  M_{\odot} \cdot yr^{-1}$ for  $q_0 = 0.5$ ($q_0 = 0.05$),
 and the faintest one has a magnitude  of $I_{AB}=26.6$ corresponding to
 a $SFR=2.2 \pm 0.3 \ (8 \pm  2) \ h^{-2} \  M_{\odot} \cdot yr^{-1}$ for  
 $q_0 = 0.5$ ($q_0 = 0.05$) 
 (the uncertainties are linked to the redshift range $3.8 \le z \le 4.4$). \\
  Assuming that the redshift range  is uniformly probed, 
  the star formation rate per unit comoving volume at $z \simeq 4.1$ 
derived from objects with $r\le 25$ varies (for cases $A$ and $B$, respectively) between 
 $10^{-2.21} - 10^{-2.14} \ (10^{-2.49} - 10^{-2.42}) ~ h^3 \cdot M_{\odot} \cdot yr^{-1} \cdot
Mpc^{-3}$ with  $q_0 = 0.5$ ($q_0 = 0.05$). For $r\le 26$, the estimates increase to  
 $10^{-2.00} - 10^{-1.82} ~ (10^{-2.28} - 10^{-2.10}) ~ h^3 \cdot
 M_{\odot} \cdot yr^{-1} \cdot Mpc^{-3}$  with  $q_0 = 0.5$ ($q_0 = 0.05$).  \\
This latter estimate is close to the value of
 $10^{-2.06\pm0.2} M_{\odot} \cdot yr^{-1}$$ \cdot Mpc^{-3}$
 computed by Madau (1997) for the HDF in the same redshift interval down to $V\le 28$ \\

%

\section{Discussion and Conclusions}
We have presented the first results of a deep multi-colour (B, V, r, I) 
field observed at ESO with the NTT telescope. A full description of the 
data reduction and analysis has been given. The final images and catalogue 
are available on the WEB site of ESO.

A comparative study with a WFPC2-HST image of the same field allows us to 
estimate the image quality of the NTTDF. Despite the broadening of the stellar 
profiles due to the effect of the seeing, the comparison of the the star/galaxy 
classification of the NTTDF objects and the one of the WFPC2 shows good agreement. 
Reliable star/galaxy separation is obtained down to I$\simeq$24 where 
the star contamination of the galaxy sample is $\simeq$1.5\%.
The comparison of the stellar content of the NTTDF with the one of the HDF 
indicates that the star density in the NTTDF is $\sim3$ times higher, and of
the 26 stars in the magnitude range $20<I<24$, 6 are  low mass stars or possible 
 brown dwarfs with a notable object having (V-I)$\simeq5.2$.

We have shown  that the galaxy number counts at faint magnitudes 
($B\ge 25$) converge to a slope $\simeq 0.32$ in all bands. The flattening 
in the blue band removes the divergence in the extragalactic background 
light (EBL) and shows that the  dominant galaxies of the EBL have B$\le$25 
(Madau 1997). As shown by Guhathakurta et al. (1990)
the contribution of high redshift galaxies ($z\ge 3$) is not more than 
10\% of the total number counts at $B\simeq$25.
In our sample the contribution to the number counts of star forming galaxies 
at $z\sim$4  is $\simeq$ 5 \%  for 24$\le r \le$ 25 and
$\simeq$ 3 \% for 25$\le r \le$26.
 If the flattening  observed  in B is due to the reddening (combination
 IGM $+$ Lyman break) of galaxies at z$\ge 3.5-4$, it cannot explain the similar 
 flattening in the V band. Thus, the high redshift star forming
 galaxies are  not a dominant  population to explain the excess of faint
 sources. \\
 At faint magnitudes, the number counts can be dominated by 
 the faint-end luminosity function (LF) at increasing redshift (with a decreasing 
 contribution of the bright galaxies - $L \ge L_{\star}$ - beyond
  z$\simeq$ 0.6-1), thus the observed slope implies that the slope of
  the faint-end LF is as steep as $\alpha\simeq -1.8$
  (by assuming a  Schechter function for the LF gives
  $dLogN/dm \propto -0.4(1+\alpha)$). Such a steep slope is supported by
 HST observations (Driver et al., 1995) where a similar value is obtained for 
 the  late-type galaxies which are the dominant population responsible for the 
 number excess. This evidence is supported  by the rapid evolution of the faint-end
 luminosity function for bluer samples observed in  the deep spectroscopic
  surveys (Lilly et al.,1996, Ellis et al., 1996) up to z$\simeq1$. \\

 The median color evolution shows a rapidly blueing with increasing magnitude
(up to $r \simeq 24.5$). 
 From $r \simeq24.5$ to $r \simeq$25.5, our data suggest that the blueing
 trend is reduced, in agreement with Smail et al. (1995), and stops beyond
  $r \le25.5$. Analyzing colors as a function of the I magnitude, a similar effect
 is observed  beyond I$\simeq$ 25. The stabilization of the blueing trends observed in the
  data is consistent with the convergence of the slope in the galaxy number-counts 
(assuming that we observe the same population in different bands).\\

 Using the approach described by Giallongo et al. (1998), we have selected a sample of high-z 
 galaxies in the range $3.8\le z\le 4.4$. The derived surface density  in our
 field is 0.8$\pm 0.1$  arcmin$^{-2}$ at $r\le 25$ and 2.0 $\pm$0.6 arcmin$^{-2}$ 
 at $r\le 26$. This sample requires spectroscopic follow-up on 8-m class telescopes
(for example the VLT with the FORS instrument).
 We have derived a lower limit for the star formation rate per unit comoving
 volume at $z\simeq 4.1$ for galaxies with $r\le 26$ of
  $10^{-1.91\pm 0.09} \ (10^{-2.19 \pm 0.09}) \  h^3 \
 M_{\odot} \cdot yr^{-1} \cdot Mpc^{-3}$
  for  $q_0 = 0.5$ ($q_0 = 0.05$).
 Our estimation  should be regarded as a lower limit due to the selection technique 
 which is biased against  Lyman $\alpha$ emission line galaxies (Hu et
 al. 1998) and  dusty galaxies undetectable with optical 
 surveys (Cimatti et al., 1998).\\

  
\begin{acknowledgements}We like to thank the other CoI of the
original  observing proposal: J.Bergeron, S.Charlot, D. Clements,
 L. da Costa, E.Egami,  B.Fort, L.Gautret,
 R.Gilmozzi, R.N.Hook, B.Leibundgut, Y.Mellier, P.Petitjean,
 A.Renzini, S.Savaglio, P.Shaver, S.Seitz and L.Yan
Special thanks are due to the ESO staff for their excellent work
on the upgrading of the NTT in the past three years, to the ESO 
astronomers who collected the observations in service mode, to Stephanie Cot\'e and Albert Zijlstra of the User 
Support Group in Garching for their efforts during the preparation and
execution of the program, to E.Hu and R.N.Hook for help on the analysis
of the HST observations of the field, to the EIS team at ESO for support 
on the data reduction and to E. Bertin for providing us an updated version 
of SExtractor and useful discussions. N. Palanque (EROS team) kindly 
provided observing time on Danish telescope for calibration control. This 
work was partially supported by the ASI contracts 95-RS-38 and by the 
Formation and Evolution of Galaxies network set up by the European Commission 
under contract ERB FMRX-CT96-086 of its TMR program. S. Arnouts has been 
supported during this work by a Marie Curie Grant Fellowship and by 
scientific visitor fellowship at ESO.   
\end{acknowledgements}

\end{document}